\documentclass{rspublic}

\usepackage{bm}

\begin{document}

\title[Maxwell and the dynamics of astrophysical discs]
{James Clerk Maxwell and the\\dynamics of astrophysical discs}

\author[G. I. Ogilvie]{Gordon I. Ogilvie}

\affiliation{Department of Applied Mathematics and Theoretical Physics,\\
University of Cambridge, Centre for Mathematical Sciences,\\
Wilberforce Road, Cambridge CB3 0WA, UK}

\label{firstpage}

\maketitle

\begin{abstract}
{Saturn's rings; astrophysical discs; kinetic theory; complex fluids}
Maxwell's investigations into the stability of Saturn's rings provide
one of the earliest analyses of the dynamics of astrophysical discs.
Current research in planetary rings extends Maxwell's kinetic theory
to treat dense granular gases of particles undergoing moderately
frequent inelastic collisions.  Rather than disrupting the rings,
local instabilities may be responsible for generating their irregular
radial structure.  Accretion discs around black holes or compact stars
consist of a plasma permeated by a tangled magnetic field and may be
compared with laboratory fluids through an analogy that connects
Maxwell's researches in electromagnetism and viscoelasticity.  A
common theme in this work is the appearance of a complex fluid with a
dynamical constitutive equation relating the stress in the medium to
the history of its deformation.
\end{abstract}

\section{Maxwell 150 years ago}

In 1856, the year that James Clerk Maxwell left Trinity College,
Cambridge to become Professor of Natural Philosophy at Marischal
College, Aberdeen, he was also writing his Adams Prize essay
\textit{On the stability of the motion of Saturn's rings}.  The prize,
named after the astronomer John Couch Adams who predicted the
existence of Neptune, was awarded every two years by St John's College
and the University of Cambridge for a mathematical essay on a
specified topic.  Among the examiners for the 1856 essay with its
planetary theme were James Challis and William Thomson, later Lord
Kelvin.  Challis was the Plumian Professor of Astronomy whose lack of
interest in Adams's prediction led him famously to fail to discover
Neptune.

Such was the perceived difficulty of the chosen subject that Maxwell's
was the only entry in the competition.  The essay topic required
Maxwell to investigate the stability of various configurations of
solid, liquid and particulate rings orbiting around Saturn.  Laplace
had already shown that a uniform solid ring would be unstable, and
suggested that the rings were solid but of non-uniform density.
Maxwell's essay is interesting partly for its use of Fourier analysis
and dispersion relations in the determination of stability criteria.
For example, he considered the behaviour of a circular ring of equally
spaced identical satellites under the destabilizing influence of their
mutual gravitational attraction.  He did this by introducing a
perturbation in the form of a displacement that was a periodic
function of the azimuthal angle $\phi$ and would depend on time
through a complex exponential function, being proportional (in a
modern notation) to $\exp(\ri m\phi+s t)$.  An algebraic dispersion
relation followed, which gave the growth rate $s$ in terms of the
wavenumber $m$.  If a solution existed with a positive real part for
any suitable value of $m$ he deduced that the configuration was
unstable and therefore untenable as a model of Saturn's rings.
Maxwell's conclusion was that ``the only system of rings which can
exist is one composed of an indefinite number of unconnected
particles, revolving around the planet with different velocities
according to their respective distances''.  The following year he
expressed himself somewhat more poetically in a letter to Thomson,
describing the rings as ``a great stratum of rubbish jostling and
jumbling round Saturn without hope of rest or agreement in itself''.

The published version of Maxwell's essay (Maxwell 1859) includes the
results of work done in 1857 after the competition was closed.  Brush,
Everitt \& Garber (1983) provide a comprehensive historical discussion
together with reprints of all the relevant documents.  A detailed
scientific critique of Maxwell's essay has been given by Cook \&
Franklin (1964).

\section{Modern view of Saturn's rings}

The modern view of Saturn's rings (\textit{e.g.}~Esposito 2006 and
references therein) benefits, of course, from space exploration,
notably the \emph{Voyager} encounters of the early 1980s and the
current \emph{Cassini} mission.  The principal rings of Saturn are the
outer A~ring and the inner B~ring, separated by the 4500-km Cassini
division.  Much of the structure in the A~ring can be attributed to
its gravitational interaction with some of the moons of Saturn.  The
narrow Encke and Keeler gaps are cleared by moonlets orbiting within
the A~ring, which also generate trailing wakes and wavy edges.  Also
seen in the A~ring are many discrete orbital resonances with more
massive moons orbiting outside the ring system; at these locations
spiral density or bending waves are launched, which propagate some
distance towards or away from Saturn before being damped.  The B~ring,
however, contains a rich irregular radial structure that cannot be
attributed to satellite resonances, and which remains one of the major
puzzles of ring dynamics (\textit{e.g.}~Tremaine 2003).

Although even \emph{Cassini} cannot resolve individual ring particles,
modern observations concur with Maxwell's conclusion regarding the
constitution of the rings.  The system consists of a dense `granular
gas' of iceballs up to several metres in diameter, the size
distribution being inferred from radio occultation experiments.  Ring
particles are in circular orbital motion in Saturn's equatorial plane,
at velocities on the order of $10~\mathrm{km}\,\mathrm{s}^{-1}$.  In
addition they have random velocities on the order of
$1~\mathrm{mm}\,\mathrm{s}^{-1}$.  The tiny ratio of $10^{-7}$ is
therefore characteristic of the orbital eccentricities and
inclinations of ring particles, and also of the angular semithickness
of the rings.

Owing to their velocity dispersion, ring particles experience
collisions that are very gentle but nevertheless significantly
inelastic because of the nature of the material.  Particles undergo a
few collisions per orbit, placing the system in an interesting but
challenging physical regime.  It is different on the one hand from the
situation in stellar dynamics, where the probability of a star
undergoing a significant scattering event with another star within one
orbit around a galaxy is typically very small.  It differs on the
other hand from gas dynamics, where molecules usually experience very
many elastic collisions within the time-scales characteristic of the
macroscopic flow and therefore establish a locally Maxwellian velocity
distribution.  The implication is that the rings do behave as a
continuous medium similar to a fluid, but a complex or non-Newtonian
one with a non-trivial rheology that needs to be explored.  The
velocity distribution can be significantly anisotropic.

Gravitational forces between ring particles are not negligible because
the escape speed from the surface of a typical particle may be
comparable to the velocity dispersion.  As well as affecting the
collisional dynamics to some extent, self-gravity as a collective
effect makes the rings thinner and is especially important in the
propagation of waves.

An important aspect of ring dynamics is the establishment of an energy
equilibrium.  Owing to Kepler's third law, the rings undergo
systematic shear, with the inner parts rotating more rapidly than the
outer.  When particles collide their relative velocity results partly
from their random velocities and partly from the systematic shear.
Velocity dispersion can therefore be generated from shear in the same
way that viscosity generates heat in a shearing fluid.  On the other
hand the inelasticity of the collisions diminishes the random
velocities.  The velocity dispersion is therefore set by the
restitution law of the material.  Experiments at UC Santa Cruz have
determined how the coefficient of restitution of ice at low
temperatures and pressures decreases with increasing collision speed
(\textit{e.g.}~Hatzes, Bridges \& Lin 1988).  An energy equilibrium in
the rings can be found when the typical collision speed is such as to
imply a critical amount of inelasticity (Goldreich \& Tremaine 1978).
If the coefficient of restitution is always too small to allow this
solution, another equilibrium may be found in which the velocity
dispersion is comparable to the shear velocity across a particle
diameter.

There are essentially two possible approaches to the theoretical study
of Saturn's rings, $N$-body simulations and statistical theories, each
of which has its merits and limitations.  Direct $N$-body simulations
(\textit{e.g.}~Salo, Schmidt \& Spahn 2001) have the character of
virtual laboratory experiments, and the accuracy of the model is
constrained mainly by our uncertainty about the physical properties of
the particles.  An obvious limitation of this approach, however, lies
in the limited scale of the simulations.  For example, the number of
metre-sized particles in Saturn's B~ring is on the order of $10^{15}$,
whereas a large-scale numerical simulation involves about $10^5$
particles.  Simulations are therefore restricted to a tiny patch of a
ring, corresponding to length-scales that are only just observable
with \textit{Cassini}; they are also limited to durations very much
shorter than the global evolutionary time-scale of the ring.  The
extreme thinness of the rings is an obstacle to global simulations,
because of the great ranges of length and time-scales involved.  In
order to understand the very direct and detailed information provided
by simulations and to apply it to phenomena on larger scales or in
different circumstances, the results need to be assimilated within a
theoretical framework of some kind.  Therefore it is important to
develop analytical models in parallel with advances in computation.

A statistical approach to planetary ring dynamics is natural in view
of the very large number of particles and the desire to formulate a
continuum description of some kind.  Nevertheless, the difficulties
involved are severe.  Such work builds on the foundations of kinetic
theory laid by Maxwell.  Indeed, the manuscripts held in Cambridge
University Library (reprinted in Brush \textit{et al.} 1983) show that
Maxwell clearly conceived the idea of the energy equilibrium in
Saturn's rings.  He also started to develop a kinetic theory of the
rings, but it proved much more difficult than the theory of gases
because of the inelastic nature of the collisions.  Their relative
infrequency also means that the velocity distribution is significantly
anisotropic and therefore non-Maxwellian.

\section{Kinetic theory of Saturn's rings}

The aim of a statistical approach is to derive a continuum mechanical
model of the rings, consisting of a set of partial differential
equations related to, but more complicated than, the Navier--Stokes
equations of fluid dynamics.  To do this one may start from Enskog's
equation governing the collisional evolution of the distribution
function $f(\bm{x},\bm{v},t)$, which gives the number density of
particles in the six-dimensional position--velocity phase space.  This
is similar to Boltzmann's equation in gas dynamics but allows for the
dense nature of the medium.  It is also necessary to modify Enskog's
equation to allow for the inelasticity of the collisions (Jenkins \&
Richman 1985; Araki \& Tremaine 1986).

In order to make analytical progress, it is supposed (in common with
most $N$-body simulations) that the particles are identical, smooth,
indestructible spheres.  Collisions are assumed to involve only two
particles at a time, and gravitational scattering or focusing is
neglected.

The description can be reduced from six dimensions to three by taking
velocity moments of Enskog's equation.  The zeroth and first moments
generate equations for the conservation of mass and momentum, while
the second moment generates a dynamical equation for the kinetic
stress or velocity dispersion tensor $\mathbf{W}$ (equation~\ref{dwdt}
below).  This can be regarded, in part, as the constitutive equation
for the complex fluid.  Rather than relating the stress
instantaneously and locally to the rate of strain, as in a Newtonian
viscous fluid, the dynamical constitutive equation accounts for the
fact that the medium has a finite `memory' of its deformation history.
To close the moment hierarchy it is necessary to neglect, or otherwise
model, the third moments that appear.  The collision term in the
second-moment equation can be evaluated by modelling the distribution
function as a triaxial Gaussian.  Experience suggests that this
relatively crude method of treating the velocity dimensions in
Enskog's equation can yield quite accurate results because of the
smoothing effect of the integral operator in the collision term.

In addition to the kinetic stress there is also a nonlocal, or
collisional, stress requiring further integrals.  This contribution
originates from the direct transfer of momentum between particles
during a collision, rather than the carriage of momentum by particles
between collisions, and can be dominant in a dense gas.

The result of this procedure is a closed system of nonlinear partial
differential equations of the form (Latter 2006)
\begin{equation}
  \frac{\partial n}{\partial t}+\bm{\nabla}\bm{\cdot}(n\bm{u})=0,
\end{equation}
\begin{equation}
  \frac{\partial}{\partial t}(mn\bm{u})+
  \bm{\nabla}\bm{\cdot}(mn\bm{u}\bm{u}+mn\mathbf{W}+\mathbf{P})=
  -mn\bm{\nabla}\Phi,
\end{equation}
\begin{equation}
  \frac{\partial\mathbf{W}}{\partial t}+
  \bm{u}\bm{\cdot}\bm{\nabla}\mathbf{W}+
  \mathbf{W}\bm{\cdot}\bm{\nabla}\bm{u}+
  (\bm{\nabla}\bm{u})^\mathrm{T}\bm{\cdot}\mathbf{W}=
  \mathbf{\dot W}_\mathrm{c},
\label{dwdt}
\end{equation}
where $m$ is the particle mass, $n$ the number density, $\bm{u}$ the
bulk velocity, $\mathbf{W}$ the velocity dispersion tensor and $\Phi$
the gravitational potential.  The collisional stress $\mathbf{P}$ and
the collisional rate of change of velocity dispersion, $\mathbf{\dot
W}_\mathrm{c}$, are complicated algebraic or numerical functions of
$n$, $\bm{\nabla}\bm{u}$ and $\mathbf{W}$.

Having set up this system of equations, one may apply it to a local
model of a differentially rotating disc known as the shearing sheet.
This uses the extreme thinness of the rings to replace the global
cylindrical geometry with a local Cartesian one in which the
systematic motion is represented as a linear shear flow in a uniformly
rotating frame of reference.  An approximate procedure of vertical
averaging is used to treat the dimension perpendicular to the ring
plane.

First, solutions may be sought that are independent of position and
time, representing the local uniform statistical equilibrium states of
the rings.  Such equilibria incorporate the energy balance mentioned
above, and predict a certain anisotropic shape for the velocity
distribution.  These solutions depend on the restitution law of the
material and on various dimensionless parameters such as the normal
optical thickness, which is an observable quantity and a measure of
the areal density of ring material.

Following Maxwell's approach, one may then study the stability of the
equilibria with respect to small disturbances with a sinusoidal
dependence on (radial) position and a complex exponential dependence
on time, being proportional to $\exp(\ri kx+s t)$.  It is possible to
derive a complicated algebraic dispersion relation (with numerically
determined coefficients) giving the growth rate $s$ as a function of
the wavenumber $k$.  If any of the solutions has a positive real part,
an instability is present.

Two types of instability that have been discussed on the basis of
viscous (\textit{i.e.} fluid dynamical) models of the rings are the
viscous instability and the viscous overstability.  The former is
favoured when the viscosity decreases with surface density, the latter
when it increases sufficiently rapidly.  In each case the instability
would occur for all wavelengths exceeding a critical value, although
the growth rate diminishes rapidly with increasing wavelength.  A
kinetic treatment of dilute rings shows that the viscous overstability
is in fact completely suppressed because of the non-Newtonian nature
of the kinetic stress (Latter \& Ogilvie 2006).  However, the
overstability is found to occur in dense rings where the collisional
stress is important, in agreement with $N$-body simulations, if the
optical thickness exceeds a critical value (Latter 2006).

In Maxwell's problem instability meant that the proposed rotating
configuration could be ruled out as a model of Saturn's rings.
However, the instabilities that occur in a kinetic model need not be
destructive to the rings.  In fact, their nonlinear evolution may lead
to a saturated state, most likely an equilibrium of a statistical
nature, involving regular or irregular structure and dynamics.  Since
the linear instabilities exist on a broad range of wavelengths, only
the nonlinear dynamics can determine the statistical distribution of
energy between the available scales.  Future work may determine
whether this behaviour can explain the rich irregular structure of
Saturn's B~ring.

\section{The broader perspective}

In astronomical terms, Saturn's rings are just one example of a
universal phenomenon: the astrophysical disc.  There are many
situations in which a disc of fluid or solid matter is found in
Keplerian orbital motion around a massive central body
(\textit{e.g.}~Frank, King \& Raine 2002).  Protoplanetary discs, such
as the `solar nebula' that once surrounded the Sun, occur as part of
the process of star formation and contain most of the angular momentum
of the cloud that collapsed to produce the star.  These discs of dusty
molecular gas last a few million years and are the sites of planet
formation.  High-energy accretion discs consisting of a fully ionized
plasma are found around black holes or compact stars in interacting
binary stars within the Galaxy, and also around much more massive
black holes at the centres of active galaxies and quasars.

Despite the very different physical constitution of these various
systems, they share common dynamical features.  In each case a shear
stress (often called a viscous stress) that develops in the
differentially rotating flow transports angular momentum outwards.  In
accretion discs this allows inward mass transport towards the central
body.  One of the challenges of accretion disc theory is to understand
and quantify the effects that give rise to this stress --- that is, to
understand the rheology of the disc.

While in planetary rings the stress arises from the kinetics of
colliding particles, in high-energy accretion discs it is believed to
result from turbulence in the presence of a magnetic field.  The
magnetorotational instability is a robust dynamical instability of a
differentially rotating flow in which the angular velocity decreases
outwards (Balbus \& Hawley 1998).  Rather than completely disrupting
the orbital motion, it generates small-scale anisotropic
magnetohydrodynamic turbulence in the disc, which transports angular
momentum outwards.  While the identification of the role of this
instability represents a great advance in the theory of accretion
discs, understanding the behaviour of the turbulent plasma is still a
major challenge that has been attempted almost exclusively with direct
numerical simulations of the magnetohydrodynamic equations in three
dimensions.

\section{A Maxwellian `mechanical model'}

In the world of nineteenth-century physics it was often seen as
desirable to develop mechanical models of new phenomena.  Maxwell
constructed a device involving ivory balls to illustrate the patterns
of displacements in his theory of Saturn's rings.  In developing his
great electromagnetic theory, he found it useful to formulate a
mechanical model of `molecular vortices' and `idle wheels' (Maxwell
1861), even though he later adopted a more abstract Lagrangian
approach (Maxwell 1865).  Here I discuss a physical and mathematical
analogy that links two quite different areas of Maxwell's work and
connects astrophysical fluid dynamics with the laboratory.

The great rivals Newton and Hooke both derived constitutive equations
for different types of medium.  While Newton proposed that the viscous
stress in a fluid is proportional to the shear rate, or rate of
strain, Hooke's law stated that the elastic stress in a solid is
proportional to the strain itself.  In his paper \emph{On the
dynamical theory of gases} (1867) Maxwell wrote down an equation that
interpolated between these descriptions and would later give rise to
the theory of viscoelasticity:
\[
  \frac{\rd\mathrm{(stress)}}{\rd t}=
  \mathrm{elasticity}\times\frac{\rd\mathrm{(strain)}}{\rd t}-
  \frac{\mathrm{stress}}{\tau}.
\]
According to Maxwell, a deformable medium has a characteristic
relaxation time $\tau$ on which molecular interactions attempt to
establish an isotropic velocity distribution and the shear stress,
which is associated with anisotropy, decays.  If a deformation is
applied that has a steady rate of strain or varies on a time-scale
long compared to $\tau$, the first term above is negligible and a
viscous response occurs with the stress being proportional to the rate
of strain as in Newton's law.  However, if the strain varies on a
time-scale short compared to $\tau$, the third term above is
negligible and an elastic response occurs with the stress being
proportional to the strain as in Hooke's law.  The reason for the
elastic response is that the rapid strain prevents the molecules from
relaxing towards an equilibrium distribution, and the stress is
therefore `frozen in' to the fluid during the deformation.  In general
Maxwell's equation allows a continuum of behaviour between viscous and
elastic.

Modern constitutive equations for viscoelastic fluids (Bird, Armstrong
\& Hassager 1987$a$) are usually expressed in a covariant tensorial
form based on the principles set out by Oldroyd (1950).  His liquid B
is one of the most widely used nonlinear models of a viscoelastic
fluid, and provides an adequate representation of a dilute solution of
a polymer of high molecular weight.  It is based on Maxwell's
viscoelastic equation above, but with a properly covariant
interpretation of the time-derivatives.  It can also be derived from
the kinetic theory of idealized long-chain polymer molecules contained
in a Newtonian solvent (Bird, Curtiss \& Armstrong 1987$b$).  If the
viscosity of the solvent is negligible the model is known as the
upper-convected Maxwell fluid.  The dimensionless number
characterizing the ratio of the relaxation time to the time-scale of
the flow is the Deborah number; when this is large, the polymeric
stress is effectively `frozen in' to the fluid.

In an electrically conducting fluid such as the fully ionized plasma
of an accretion disc around a black hole, the magnetic field $\bm{B}$
affects the dynamics through the bulk Lorentz force (e.g. Roberts
1967).  This can be represented in terms of the Maxwell
electromagnetic stress tensor,
\begin{equation}
  \mathbf{M}=\frac{\bm{B}\bm{B}}{\mu_0}-\frac{B^2}{2\mu_0}\mathbf{1},
\end{equation}
the two parts of which correspond to a tension in the field lines and
an isotropic magnetic pressure (for non-relativistic flows, the
electric field makes a negligible contribution to the stress).  It is
well known that, in a perfectly conducting fluid, the magnetic field
is `frozen in' to the fluid, in the sense that magnetic field lines
can be identified with material lines (Alfv\'en 1950).  Even in a
fluid of finite conductivity, the magnetic field is effectively
`frozen in' for motions of sufficiently short time-scale, or
sufficiently large length-scale, corresponding to a large magnetic
Reynolds number.  It follows that the Maxwell stress is also `frozen
in' to the fluid in a certain sense.

Ogilvie \& Proctor (2003) have shown that there is a physical and
mathematical similarity between the dynamics of viscoelastic and
magnetohydrodynamic fluids.  Either the polymer molecules or the
magnetic field lines are advected and distorted by the flow and exert
a tension force on it.  A formal mathematical analogy can be drawn
between the Oldroyd-B fluid in the limit of large Deborah number and a
magnetized plasma in the limit of large magnetic Reynolds number.
Either the polymeric tension or the magnetic tension in this limit
satisfies the equation
\begin{equation}
  \frac{\partial\mathbf{T}}{\partial t}+
  \bm{u}\bm{\cdot}\bm{\nabla}\mathbf{T}-
  \mathbf{T}\bm{\cdot}\bm{\nabla}\bm{u}-
  (\bm{\nabla}\bm{u})^\mathrm{T}\bm{\cdot}\mathbf{T}=0,
\end{equation}
which states that the stress is `frozen in' to the flow.  This
equation is also closely related to equation~(\ref{dwdt}) for the
kinetic stress in a planetary ring, although there is an important
change of sign in two of the terms.  The left-hand sides of these
equations are both Lie derivatives, one for a contravariant tensor and
the other for a covariant tensor.

Ogilvie \& Proctor (2003) also showed that the magnetorotational
instability, which, as mentioned above, is of primary importance in
the theory of accretion discs, has a direct analogue in a viscoelastic
fluid.  Experiments have been carried out on the stability of
viscoelastic Couette--Taylor flow between differentially rotating
concentric cylinders since the 1960s.  However, they appear to have
neglected the regime in which the magnetorotational instability would
appearmost clearly; this requires that both cylinders rotate, but in
such a way that the specific angular momentum increases outwards while
the angular velocity decreases, as in a Keplerian disc.  It is likely
that an experiment using a dilute polymer solution could quite easily
demonstrate the existence of this astrophysically important
instability and determine its nonlinear evolution.  In contrast,
experimental efforts to demonstrate the instability in its
magnetohydrodynamic version still face considerable technical
difficulties (Goodman \& Ji 2002).

\section{Conclusion}

Planetary rings provide the only examples of the universal phenomenon
of astrophysical discs that can be observed with exquisite resolution.
Investigations of the stability of Saturn's rings 150 years after
Maxwell's still hold interest as they may explain the rich irregular
structure in the B~ring.  Understanding the behaviour of planetary
rings in detail has been difficult because of the complex physical
nature of the medium: a non-Newtonian fluid consisting of a dense
granular gas of particles undergoing inelastic and only moderately
frequent collisions.  Current research builds on the kinetic theories
developed by Maxwell and others.

Other astrophysical discs also possess a complicated rheology owing to
the presence of magnetohydrodynamic turbulence in a fully ionized
plasma.  A simple analogy with polymeric liquids may provide a way to
understand aspects of their behaviour and even to study it
experimentally.  In all of this work the theory rests on the firm
foundations established by Maxwell's pioneering researches in
dynamics, kinetic theory, electromagnetism and viscoelasticity.

\begin{acknowledgements}
I am grateful to the organisers of the meeting, John Reid and Charles
Wang, for their invitation and their hospitality in Aberdeen.  I also
thank Henrik Latter for discussions.  This research was supported in
part by the Leverhulme Trust.
\end{acknowledgements}

\end{document}